\documentclass[preprintnumbers,showkeys,superscriptaddress,showpacs,
nofootinbib,byrevtex,fleqn]{revtex4}   
\usepackage{amsmath,amsfonts,amssymb,amscd,amsxtra,amsthm}
\usepackage{graphicx}
\usepackage{bm}
\usepackage{epstopdf}
\begin{document}
\preprint{CYCU-HEP-11-01}
\title{$\Lambda(1520)$ photoprodcution with Regge contribution}
\author{Seung-il Nam}
\affiliation{Research Institute of Basic Sciences, Korea Aerospace
University, Goyang, 412-791, Korea}  
\author{Chung-Wen Kao}
\affiliation{Department of Physics, Chung-Yuan Christian University, Chung-Li 32023, Taiwan}  
\author{Byung-Geel Yu}
\affiliation{Research Institute of Basic Sciences, Korea Aerospace
University, Goyang, 412-791, Korea}  
\begin{abstract}
In this talk, we report our recent progresses on the $\Lambda(1520)$ photoproduction using the effective Lagrangian approach. In addition to the tree-level Born diagrams, we take into account the Regge-trajectories for the possible strange-meson exchanges in the $t$ channel. 
We compute the angular and energy dependences of the production process, including polarization observables, such as the photon-beam asymmetry and the polarization-transfer coefficients, resulting in good qualitative agreement with current experimental data. We also compute the $K^{-}$-angle distribution function in the Gottfried-Jackson frame, using the polarization-transfer coefficients in the $z$ direction. 
\end{abstract}

\maketitle
\section{Introduction}
The photoproduction of hyperons off the nucleon target is important in hadron physics since it reveals the strangeness-related interaction structures of hadrons. Many experimental collaborations, such as CLAS at Jafferson laboratory~\cite{McNabb:2003nf,Bradford:2005pt,Bradford:2006ba}, LEPS at SPring-8~\cite{Niiyama:2009zz,Kohri:2009xe,Hicks:2008yn,Muramatsu:2009zp}, etc, have performed energetic research activities for the photoproductions. Up to the resonance region $\sqrt{s}\lesssim3$ GeV, a simple hadronic model including the tree-level diagrams has successfully explained the experimental data. However, as energy increases, this simple model needs to be extended to accommodate the effects from the interactions at the quark level. To reach this goal, mesonic Regge trajectories, corresponding to all the meson exchanges with the same quantum numbers but different spins in $t$ channel at tree level, were employed~\cite{Corthals:2005ce,Corthals:2006nz,Ozaki:2009wp}. 

In the present report, we investigate the $\Lambda(1520,3/2^{-})\equiv\Lambda^{*}$ photoproduction off the proton target, $\gamma p\to K^{+}\Lambda^{*}$,
beyond the resonance region with the extended model including the original hadronic model and the interpolated Regge contributions. As shown in Ref.~\cite{Nam:2005uq}, up to the resonance region, this production process is largely dominated by the contact-term contribution. Note that this interesting feature supported by the experiments~\cite{Muramatsu:2009zp,Nakano} is a consequence of gauge invariance in a certain description for spin-$3/2$ fermions. For instance, according to it, 1) one can expect a significant difference in the production strengths. We present the energy and angular dependences, photon-beam asymmetry, and polarization-transfer coefficients of the production process. Furthermore the $K^{-}$-angle distributions function, $\mathcal{F}_{K^{-}}$ in the Gottfried-Jackson frame using the polarization transfer coefficients $C_{z,1/2}$ and $C_{z,3/2}$
are also computed. Due to the limited volume for the conference proceeding, we would like to focus on the $K^{-}$-angle distribution function here. More detailed results for various physical observables can be found in Refs.~\cite{Nam:2005uq,Nam:2010au,Nam:2009cv}.

\subsection{$K^{-}$-angle  distribution function}
One of our research focusses in the present report is the $K^{-}$-angle distribution function~\cite{Barber:1980zv,Muramatsu:2009zp}, which is the angle distribution of $K^{-}$, decayed from $\Lambda^{*}$ ($\Lambda^{*}\to K^{-}p$), in the $t$-channel helicity frame, i.e. the Gottfried-Jackson frame~\cite{Schilling:1969um}. From this function, one can see which meson-exchange contribution dominates the production process. According to a simple spin statistics, the function becomes $\sin^{2}\theta_{K^{-}}$ for $\Lambda^{*}$ in $S_{z}=\pm3/2$, whereas $\frac{1}{3}+\cos^{2}\theta_{K^{-}}$ for $\Lambda^{*}$ in $S_{z}=\pm1/2$:
\begin{equation}
\label{eq:DF}
\mathcal{F}_{K^{-}}
=A\sin^{2}\theta_{K^{-}}+B\left(\frac{1}{3}+\cos^{2}\theta_{K^{-}}\right),
\end{equation}
where we have used a notation $\mathcal{F}_{K^{-}}$ indicating the distribution function for convenience. The coefficients $A$ and $B$ stand for the strength of each spin state of $\Lambda^{*}$, satisfying the normalization $A+B=1$. 

Now, we want to provide theoretical estimations on $\mathcal{F}_{K^{-}}$. Taking into account the outgoing kaon ($K^{+}$) does carry zero spin, all the photon helicity will be transffered to $\Lambda^{*}$ through the exchanging particle in $t$-channel, Hence, it is natural to think that the polarization-transfer coefficients in the $z$ direction should relate to the strength coefficients $A$ and $B$. Hence, we can write $A$ and $B$ in terms of the polarization transfer coefficients $C_{z,1/2}$ and $C_{z,3/2}$ as follows:
\begin{equation}
\label{eq:AAA}
A=\frac{C_{z,3/2}}{C_{z,1/2}+C_{z,3/2}},\,\,\,\,
B=\frac{C_{z,1/2}}{C_{z,1/2}+C_{z,3/2}},
\end{equation}
which satisfy the normalization condition. In other words, $A$ denotes the strength that $\Lambda^{*}$ is in its $S_{z}=\pm3/2$ state, and $B$ for $S_{z}=\pm1/2$ similarly.

\begin{figure}[ht]
\includegraphics[width=16cm]{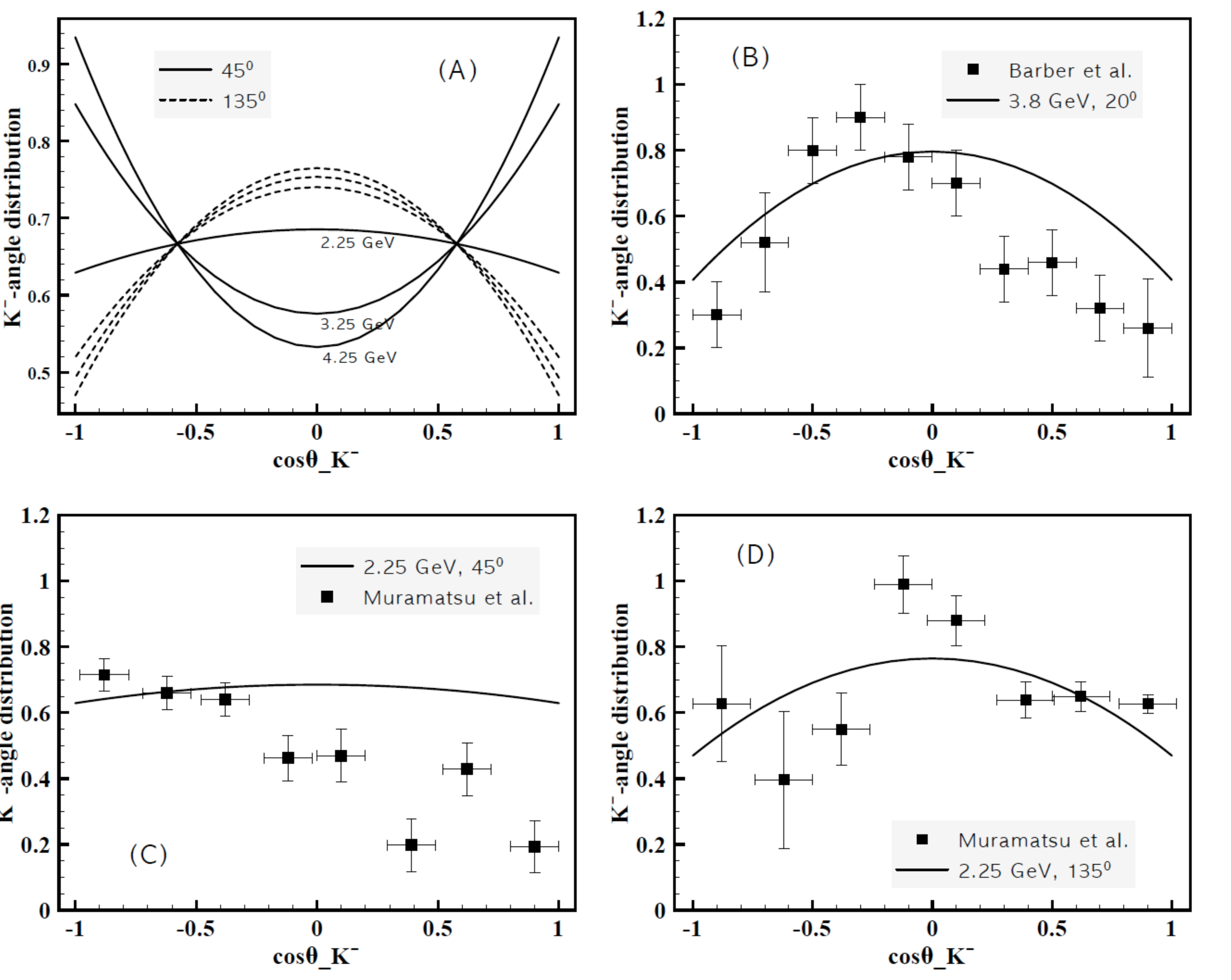}
\caption{$\mathcal{F}_{K^{-}}$ as a function of $\cos\theta_{K^{-}}$ for $E_{\gamma}=2.25$ GeV, $3.25$ GeV, and $4.25$ GeV at $\theta=45^{\circ}$ and $135^{\circ}$ in (A). In (B), we compare the numerical result for $E_{\gamma}=2.25$ geV and $\theta=30^{\circ}$ with the experimental data taken from Ref.~\cite{Barber:1980zv}. Similarly, we show the comparisons in (C) and (D) for $\theta=45^{\circ}$ and $135^{\circ}$, respectively, with Ref.~\cite{Muramatsu:2009zp}. See the text for details.}
\label{FIG13}
\end{figure}

In (A) of Figure~\ref{FIG13}, we draw $\mathcal{F}_{K^{-}}$ as a function of $\cos\theta_{K^{-}}$ for $E_{\gamma}=2.25$ GeV, $3.25$ GeV, and $4.25$ GeV at $\theta=45^{\circ}$ and $135^{\circ}$. As seen in the figure, in the backward-scatreing region represented by $\theta=135^{\circ}$, the curves for $\mathcal{F}_{K^{-}}$ are similar to each other $\sim\sin^{2}\theta_{K^{-}}$. On the contrary, they are different considerably for the forward-scattering region represented by $\theta=45^{\circ}$, depending on $E_{\gamma}$. The curves, which are proportional to $\sin^{2}\theta_{K^{-}}$ or $\frac{1}{3}+\cos^{2}\theta_{K^{-}}$, are mixed, and the portion of each contribution depends on $E_{\gamma}$. In (B), we compare the numerical result for $E_{\gamma}=3.8$ GeV at $\theta=20^{\circ}$ with the experimental data taken from Ref.~\cite{Barber:1980zv} for $E_{\gamma}=(2.8\sim4.8)$ GeV and $\theta=(20\sim40)^{\circ}$. We normalize the experimental data with the numerical result by matching them at $\theta_{K^{-}}=90^{\circ}$ approximately. The theory and experiment are in a qualitative agreement, $\mathcal{F}_{K^{-}}\propto\sin^{2}\theta_{K^{-}}$.  Although we did not show explicitly, theoretical result for $\mathcal{F}_{K^{-}}$ changes drastically around  $\theta=25^{\circ}$. At $\theta\approx30^{\circ}$, the curve turns into $\sim\frac{1}{3}+\cos^{2}\theta_{K^{-}}$.

Similarly, we compare them in (C) and (D) for $\theta=45^{\circ}$ and $135^{\circ}$, respectively, for $E_{\gamma}=2.25$ GeV with Ref.~\cite{Muramatsu:2009zp} for $E_{\gamma}=(1.75\sim2.4)$ GeV and $\theta=(0\sim180)^{\circ}$. Again, we normalized the experimental data to the numerical result for the backward-scattering region (D) as done above. Then, we used the same normalization for the forward-scattering region (C). As shown in (C), the experiment and theory starts to deviate from each other beyond $\cos\theta_{K^{-}}\approx-0.2$. In Ref.~\cite{Muramatsu:2009zp}, it was argued that there can be a small destructive interference caused by the $K^{*}$-exchange contribution to explain the experimental data shown in (C). However, this is unlikely, since that of $K^{*}$ exchange only gives negligible effect on $C_{z,1/2}$ and $C_{z,3/2}$~\cite{Nam:2009cv}. Hence, we consider the large deviation in (C) may come from the interference between $\Lambda^{*}$ and other hyperon contributions, which are not taken into account in the  present work. As for the backward-scattering region, $\mathcal{F}_{K^{-}}$ shows a curve $\sim\sin^{2}\theta_{K^{-}}$, and the experimental data behaves similarly. Although we have not considered the interference, for these specific angles, present theoretical  estimations on $A$ are very similar to those given in Ref.~\cite{Muramatsu:2009zp}.

\section{Summary and outlook}
We have investigated the $\Lambda(1520)$ photorproduciton. Using the tree-level Born approximation plus the Regge contributions, we could reproduce the experimental data qualitatively very well. The interpolation between the Feynman and Regge propagators turns out to be crucial to describe the wide energy-range data. Our theoretical estimations for the $K^{-}$-angle  distribution function can be a useful guide for the future experiments. Although the nucleon resonance contributions seem negligible for relatively wide-angle regions, hyperon resonances may play an important role, since the backward scattering regions have not been reproduced well.  Moreover, the asymmetric curve for the $K^{-}$-angle  distribution at $(E_\gamma,\theta_{K^+})=(2.25\,\mathrm{GeV},45^\circ)$ in the experimental data, being different from the numerical result, must be addressed carefully in the further studies. Related works will appear elsewhere. 
\section*{Acknowledgments}
The present report was prepared as a proceeding for the international concference BARYONS'10, $7\sim11$ December 2010, Osaka, Japan. The authors appreciate the hospitality during their attending the conference. They are also grateful to A.~Hosaka, W.~C.~Chang and H.~-Ch.~Kim for fruitful discussions. The works of S.i.N. and B.G.Y. were supported by the grant NRF-2010-0013279 from National Research Foundation (NRF) of Korea. S.i.N. was also partially supported by the grant NSC 96-2112-M-033-003-MY3 from National Science Council (NSC) of Taiwan. The work of C.W.K. was supported by the grant NSC 99-2112-M-033-004-MY3 from National Science Council (NSC) of Taiwan.



\begin{thebibliography}{99}

\bibitem{McNabb:2003nf}
  J.~W.~C.~McNabb {\it et al.}  [The CLAS Collaboration],
  Phys.\ Rev.\  C {\bf 69}, 042201 (2004).
\bibitem{Bradford:2005pt}
  R.~Bradford {\it et al.}  [CLAS Collaboration],
  Phys.\ Rev.\  C {\bf 73}, 035202 (2006).
\bibitem{Bradford:2006ba}
  R.~Bradford {\it et al.}  [CLAS Collaboration],
  Phys.\ Rev.\  C {\bf 75}, 035205 (2007).
\bibitem{Niiyama:2009zz}
  M.~Niiyama  [LEPS TPC Collaboration],
  Nucl.\ Phys.\  A {\bf 827}, 261C (2009).
\bibitem{Kohri:2009xe}
  H.~Kohri {\it et al.}  [LEPS Collaboration],
  arXiv:0906.0197 [hep-ex].
\bibitem{Hicks:2008yn}
  K.~Hicks {\it et al.}  [LEPS Collaboration],
  Phys.\ Rev.\ Lett.\  {\bf 102}, 012501 (2009).
\bibitem{Muramatsu:2009zp}
  N.~Muramatsu {\it et al.},
  Phys.\ Rev.\ Lett.\  {\bf 103}, 012001 (2009).
\bibitem{Corthals:2005ce}
  T.~Corthals, J.~Ryckebusch and T.~Van Cauteren,
  Phys.\ Rev.\  C {\bf 73}, 045207 (2006).
\bibitem{Corthals:2006nz}
  T.~Corthals, D.~G.~Ireland, T.~Van Cauteren and J.~Ryckebusch,
  Phys.\ Rev.\  C {\bf 75}, 045204 (2007).
\bibitem{Ozaki:2009wp}
  S.~Ozaki, H.~Nagahiro and A.~Hosaka,
  arXiv:0910.0384 [hep ph].
\bibitem{Nam:2005uq}
  S.~i.~Nam, A.~Hosaka and H.~-Ch.~Kim,
  Phys.\ Rev.\  D {\bf 71}, 114012 (2005).
\bibitem{Nakano}
Private communications with T.~Nakano for LEPS collaboration.
\bibitem{Nam:2010au}
  S.~i.~Nam and C.~W.~Kao,
  Phys.\ Rev.\  C {\bf 81}, 055206 (2010)
\bibitem{Nam:2009cv}
  S.~i.~Nam,
  Phys.\ Rev.\  C {\bf 81}, 015201 (2010)
\bibitem{Barber:1980zv}
  D.~P.~Barber {\it et al.},
  Z.\ Phys.\  C {\bf 7}, 17 (1980).
\bibitem{Schilling:1969um}
  K.~Schilling, P.~Seyboth and G.~E.~Wolf,
  Nucl.\ Phys.\  B {\bf 15}, 397 (1970)
  [Erratum-ibid.\  B {\bf 18}, 332 (1970)].
\end{thebibliography}
\end{document}